\documentclass{ieeeaccess}
\usepackage{placeins}
\usepackage{cite}
\usepackage{amsmath,amssymb,amsfonts}
\usepackage{graphicx}
\usepackage{textcomp}
\usepackage{booktabs}
\usepackage{algpseudocode}
\usepackage{array}
\usepackage{longtable}
\usepackage{booktabs}

\usepackage{bm}
\makeatletter
\AtBeginDocument{\DeclareMathVersion{bold}
\SetSymbolFont{operators}{bold}{T1}{times}{b}{n}
\SetSymbolFont{NewLetters}{bold}{T1}{times}{b}{it}
\SetMathAlphabet{\mathrm}{bold}{T1}{times}{b}{n}
\SetMathAlphabet{\mathit}{bold}{T1}{times}{b}{it}
\SetMathAlphabet{\mathbf}{bold}{T1}{times}{b}{n}
\SetMathAlphabet{\mathtt}{bold}{OT1}{pcr}{b}{n}
\SetSymbolFont{symbols}{bold}{OMS}{cmsy}{b}{n}
\renewcommand\boldmath{\@nomath\boldmath\mathversion{bold}}}
\makeatother

\def\BibTeX{{\rm B\kern-.05em{\sc i\kern-.025em b}\kern-.08em
    T\kern-.1667em\lower.7ex\hbox{E}\kern-.125emX}}

\begin{document}
\history{Date of publication xxxx 00, 0000, date of current version xxxx 00, 0000.}
\doi{10.1109/ACCESS.2024.DOI Number}

\title{LEVERAGING BI-FOCAL PERSPECTIVES AND GRANULAR FEATURE INTEGRATION FOR ACCURATE RELIABLE EARLY ALZHEIMER’S DETECTION }
\author{\uppercase{Shravan Venkatraman}\authorrefmark{1},
\uppercase{Pandiyaraju V}\authorrefmark{2},
\uppercase{Abeshek A}\authorrefmark{3},
\uppercase{Pavan Kumar S}\authorrefmark{4},
\uppercase{Aravintakshan S A}\authorrefmark{5}}

\address[1-5]{School of Computer Science and Engineering, Vellore Institute of Technology, Chennai, Tamil Nadu, India \\
(e-mails: \{pandiyaraju.v@vit.ac.in, shravan.venkatraman18@gmail.com, abeshek.a@gmail.com, 
s.pavankumar2003@gmail.com, aravintcs176@gmail.com\})}



\corresp{Corresponding author: Pandiyaraju V (e-mail: pandiyaraju.v@vit.ac.in)}

\begin{abstract}
Being the most commonly known neurodegeneration, Alzheimer’s Disease (AD) is annually diagnosed in millions of patients. The present medical scenario still finds the exact diagnosis and classification of AD through neuroimaging data as a challenging task. Traditional CNNs can extract a good amount of low-level information in an image while failing to extract high-level minuscule particles, which is a significant challenge in detecting AD from MRI scans. To overcome this, we propose a novel Granular Feature Integration method to combine information extraction at different scales along with an efficient information flow, enabling the model to capture both broad and fine-grained features simultaneously. We also propose a Bi-Focal Perspective mechanism to highlight the subtle neurofibrillary tangles and amyloid plaques in the MRI scans, ensuring that critical pathological markers are accurately identified. Our model achieved an F1-Score of 99.31\%, precision of 99.24\%, and recall of 99.51\%. These scores prove that our model is significantly better than the state-of-the-art (SOTA) CNNs in existence.
\end{abstract}


\titlepgskip=-21pt

\maketitle

\section{Introduction}
\label{sec:introduction}
\PARstart{C}{haracterized} by chronic neurodegeneration, Alzheimer’s disease causes systematic brain neuron deterioration. It affects the cognitive abilities as well as reduces the individual’s ability to perform simple tasks thereby turning out to be the biggest cause of dementia in older people \cite{suk2014hierarchical}. An individual diagnosed with AD suffers from memory loss, problems in speaking, inability to understand, etc. Apart from these symptoms, AD also results in the formation of abnormal masses (amyloid plaques) and knotted strings of fibers known as neurofibrillary tangles of the brain. Furthermore, disconnection of neurons is observed as well constituting to the hallmark features of AD \cite{jo2020deep}. The disease primarily targets regions of the brain responsible for cognition, memory, and communication. The hippocampus and the entorhinal cortex of the brain, known for their responsibility to create a memory, are among the earliest to be targeted. The underlying cause of AD remains is still not well known, it is speculated that it might be an amalgamation of genetic factors along with lifestyle and environmental conditions.

\begin{figure*}[h]
  \centering
  \includegraphics[width=0.7\textwidth]{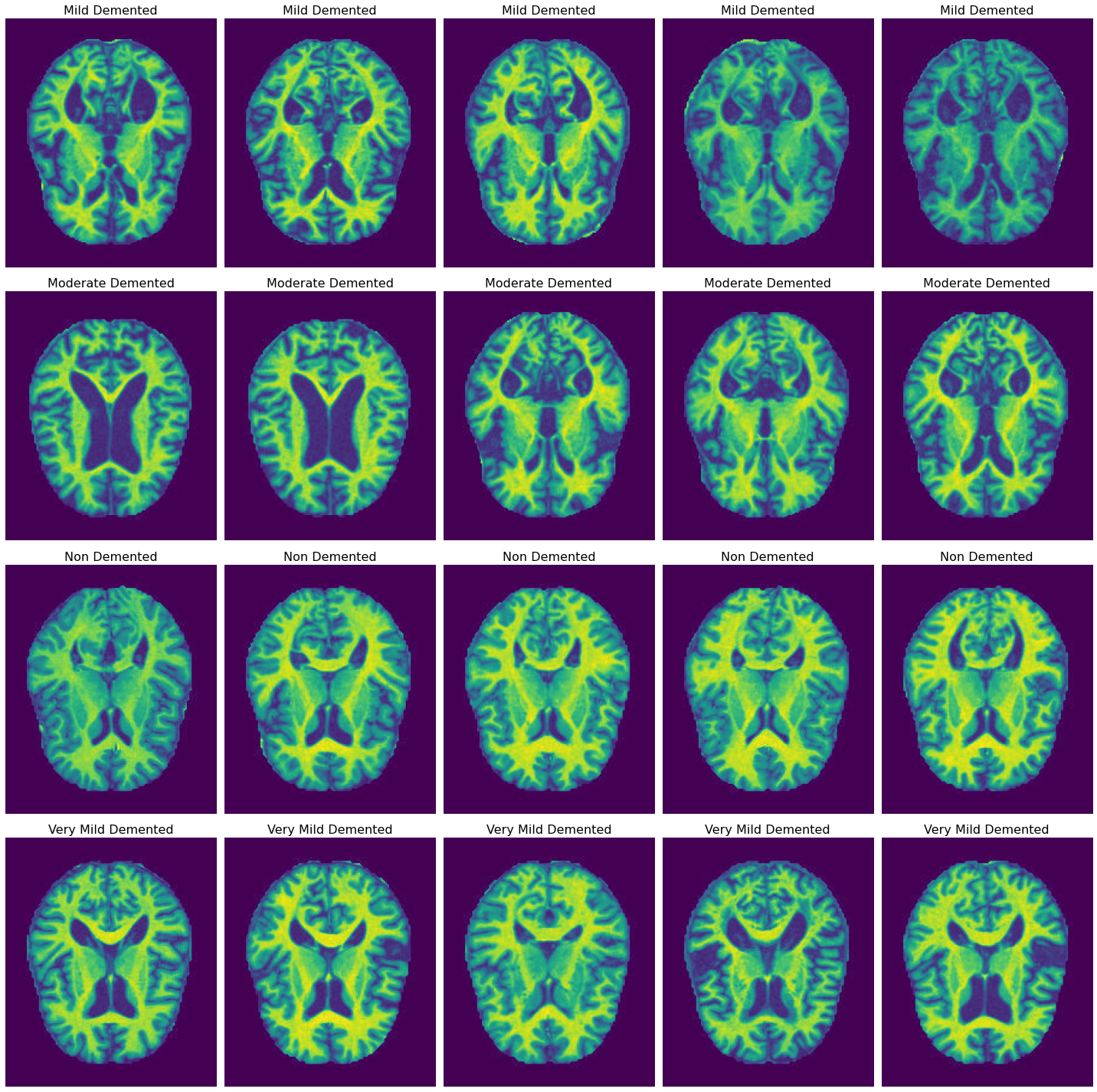}
  \caption{Sample Images of MRI Scans from the OASIS Dataset Representing Each Class of AD}
  \label{fig:figRaw}
\end{figure*}

Age is considered to be the primary risk factor since majority of positive cases are recorded in patients aged over 65 years. Genetics play a crucial role, since a family history of AD also increases the risk. Diagnosis of AD is a tedious task involving exhaustive procedures like evaluating the family history, physical examination, cognitive and neurological tests. MRI scans with doctors’ assistance are given great importance to diagnose AD. These MRI scans provide an accurate image of the brain including the description of their shape and size. Comparing them with the history of the patient, memory decline and other cognitive functions are considered to be the primary symptoms of dementia \cite{liu2014early}. In this work, we aim to contribute to the advancing AD detection by development of a state-of-the-art mechanism through our proposed method, thereby efficiently improving the quality of medical diagnosis by adopting Deep Learning (DL) algorithms.

Complex DL algorithms have a major influence in detecting AD due to their capability to analyse wide range of data. But the underlying reason for utilization of DL algorithms is because of its ability to process and interpret MRI scans to identify patterns of brain degeneration. Convolutional Neural Networks (CNNs), in particular can recognize such intricate patterns in image data which in association with advanced algorithms enable early diagnosis of AD and contribute in assisting with slowing down its progression towards dementia \cite{spasov2019parameter}. DL models provide an automatic analysis of MRI scans from patients, improving diagnostic efficiency while yielding promising results in AD identification \cite{jo2019deep}.

The DL model, BFPCNN, introduced in this research article, utilizes Bi-Focal Perspectives and integrates Granular Features for robust extraction of high-level and low-level features to help healthcare professionals to assist healthcare professionals to identify abnormalities in the brain in an efficient manner while consuming lesser duration at the same time. We utilized the open-access series of imaging studies, OASIS-1 dataset \cite{marcus2007open}, which includes cross-sectional MRI data acquired from nondemented young, demented older adults, and the middle-aged. Sample images from each class of the dataset is shown in Figure 1. Our research is then compared with that previously conducted and other high-performance CNN models. Consequently, our results show that our model outperforms previous works along with SOTA CNNs, highly accurately extracting features and classifying the MRI scans.

The contributions of this research are as follows:
\begin{itemize}
    \item \textbf{Granular Feature Integration for Multi-Scale Feature Extraction:}  
    The proposed BFPCNN introduces a Granular Feature Integration mechanism, effectively addressing the limitations of traditional CNNs that struggle to extract fine-grained, high-level details necessary for Alzheimer's diagnosis. This method ensures that the model captures both broad structural patterns and fine-grained pathological markers, such as subtle brain abnormalities, overcoming the traditional trade-off between local and global feature extraction.

    \item \textbf{Bi-Focal Perspective Mechanism in BFPCNN for Highlighting Key Biomarkers:}  
    The proposed Bi-Focal Perspective mechanism specifically enhances the identification of critical Alzheimer's biomarkers, such as neurofibrillary tangles and amyloid plaques, which are often missed by conventional models. By focusing on these subtle features, the proposed model directly addresses the challenge of detecting intricate pathological indicators in MRI scans, improving diagnostic precision.
\end{itemize}

\section{Related Works}

Helaly et al. \cite{helaly2021deep} showcase the potential of deep learning, CNNs, in the early-stage detection of AD via biomedical images. Their study is based on multi-class as well as binary classification in order to classify between different stages of AD; it supports proper diagnosis and further treatment plan. They enhance the detection with sparse data by fine-tuning models such as VGG19 through transfer learning. On the other hand, remote AD screening proposal eradicates the limitations of COVID-19 and facilitates secure patient assessment. They use key evaluation metrics like specificity, accuracy, and sensitivity to allow one to assess model performance and drive underlining clinical efficiency. Venugopalan et al. \cite{venugopalan2021multimodal} research on the detection of AD by overcoming the limitations based on single-modal data and, in turn, advocate for the integration of multi-data that better improves the analysis of AD staging. The authors introduced a novel approach where MRI images, SNPs, and clinical test data were combined using DL techniques, including stacked denoising auto-encoders for clinical and genetic data and 3D-CNN for imaging. Cases were classified according to the three AD, MCI, and control groups, and a distinct kind of clustering and perturbation analysis pointed out the top features learned by deep models. Tested on ADNI data, this multi-modal approach outperformed standard models like random forests and SVMs, even bringing about distinctions in brain areas such as the hippocampus and the amygdala.

Ahila et al. \cite{a2022evaluation} presented a study addressing the challenge imposed upon modern healthcare through AD. This they developed by creating an advanced CAD system built with CNNs trained on Positron Emission Tomography scans to enhance the system with diagnostic accuracy. Their model was trained on a set of 855 patient samples from the AD Neuroimaging Initiative (ADNI) database and tested for remarkable performance metrics with a sensitivity of 96\%, specificity of 94\%, and accuracy of 96\%. Puente-Castro et al. \cite{puente2020automatic} conceptualized early detection of AD and suggested a system that detects the existence of AD in MRI scans. They made use of transfer learning (TL) by using data from the ADNI dataset and OASIS dataset. Their model was trained to detect AD from sagittal MRIs with results that could be comparable to those gathered from horizontal-plane MRIs. This research shows the potential for early AD detection and monetary savings within DL models.

Zhang et al. \cite{zhang2022single} discussed detection regarding AD and MCI:. They have proposed a novel DL approach that is CNN and a Two-stage Random RandAugment (TRRA) for data augmentation. It has included one method which generates heat maps using Grad-CAM++. This research provides solutions towards CAD methods for AD and MCI detection to problems like data leakage, overfitting, and an opaque diagnosis. Liu et al. \cite{liu2020new} have offered a new ML method to detect the symptoms of AD. By incorporating several techniques of ML, they are working to make the diagnosis of AD very efficient as well as accurate for a better outcome in diagnosis. The research is a significant contribution to the field of medical diagnosis and serves to introduce potential improvements in AD diagnosis with potential far-reaching implications for both clinical practice and patient care. Chang et al., \cite{chang2021machine} analyzed various ML techniques along with biomarkers for AD diagnosis. In their review, they have particularly underlined the useful applicability of AI and ML tools to accurately diagnose AD. In addition to the classical biomarkers such as A$\beta$42 and tau proteins, the paper references novel potential biomarkers implicated in various mechanisms that lead to AD pathology: neuronal injury, synaptic dysfunction, and neuroinflammation. The authors also mention that ML algorithms may be applicable to the area, pointing to a necessity for fast, inexpensive, and novel diagnostic tools. 

Ebrahimighahnavieh et al. \cite{ebrahimighahnavieh2020deep} developed a literature review on various DL techniques for AD detection. During the review, they highlighted the need for reliable and efficient diagnostic tools for the detection of AD. Following this review, the authors make a significant contribution to the DL techniques, unveiling the key features, methodologies, and the difficulties encountered while providing valuable perspectives for future studies in this area. Ji et al. \cite{ji2019early} emphasized the need to detect AD earlier for proper intervention in slowing the disease progression. The authors applied DL algorithms for the task of pattern classification although the ML methods are highly dependent on manually extracted features and complex architectures. They proposed an ensemble model based on ConvNets. The authors demonstrated the DL techniques' potential within the domain of AD diagnosis. A sophisticated approach for the detection of AD using multi-modal ML was presented by Khan et al. \cite{khan2020improved}. Here, they proposed a five-stage ML pipeline that incorporates data transformation and feature selection along with a random forest classifier. The study contributed to understanding the possibility of automating diagnosis for clinical applications. 

Orouskhani et al. \cite{orouskhani2020alzheimers} introduced a novel conditional deep triplet network for AD detection from structural MRI. They addressed this problem from having a few samples  the dataset by using deep metric learning techniques. Their conditional loss function-based model, designed on top of VGG16, demonstrated superior performance compared to existing models. Saleh et al. \cite{saleh2020alzheimers} propose a class model for AD classification by using DenseNet with a healthcare decision support embedded into it. They use TL techniques to enhance their performance and demonstrate the generalization ability of the DenseNet model. Dua et al. \cite{dua2020cnnrnnlstm} propounded a new integration method of LSTM-CNN-RNN in the detection of AD from MRI scans. They tried to classify the dementia levels efficiently as well as accurately for the early onset of medical intervention. This incorporates the utilization of CNNs in feature extraction and LSTM and RNN in sequence learning. They show the use of DL algorithms for enhancing the diagnosis of AD using ensemble techniques. 

Saratxaga et al. \cite{saratxaga2020mri} developed a DL-based method to predict AD from MRI images. They discussed how early diagnosis for timely intervention would be of extreme importance for AD. Their proposed method is a combined DL and image processing technique that depicts some remarkable improvements in the diagnosis of AD. They achieved BAC of 0.93 in automated diagnosis and 0.88 for the classification of disease stages. Liu et al \cite{liu2020alzheimers} proposed a novel approach that utilized depthwise separable CNNs (DSC) for the detection of AD from MRI scans. DSC reduces the complexity of the model and decreases the overhead of computations without compromising the performance. They could show that TL is feasible to be used on the ground with AlexNet and GoogLeNet in practice within mobile embedded systems.

The reviewed studies discussed various DL and ML approaches that have improved early detection and classification of AD by using biomedical imaging and multi-modal data. Transfer learning has been one of the most applied methods, using a pre-trained model: VGG19, DenseNet, GoogLeNet, etc, to leverage knowledge on sparse datasets. Features extraction in CNNs is also applied. Further reduction in the complexity of computations as compared to accuracy is achieved by depthwise separable CNNs. Ensemble models as a combination of CNNs, RNNs and LSTMs are also formed for sequential learning and multi-stage classification. Different kinds of data such as MRI scan data, SNPs, and clinical test data, provide multi-modal approaches improving the diagnostic power with capture of the various dimensions of the disease. Techniques like Grad-CAM++ and perturbation analysis have been used to apply interpretability to find important features in biomedical images. Techniques such as two-stage random RandAugment have been used for overcoming the problem of overfitting. Advanced architectures such as conditional deep triplet networks solved the problem of scarcity of data through metric learning. New pipelines that apply feature selection and transformation are being used for enhancing ML model performance. Collectively, these approaches advance the state of the art in terms of accuracy, scalability, and interpretability of AD detection systems toward effective clinical applications.

\section{Proposed Work}

\subsection{Dataset Exploration}

\begin{figure*}[h]
  \centering
  \includegraphics[width=0.65\textwidth]{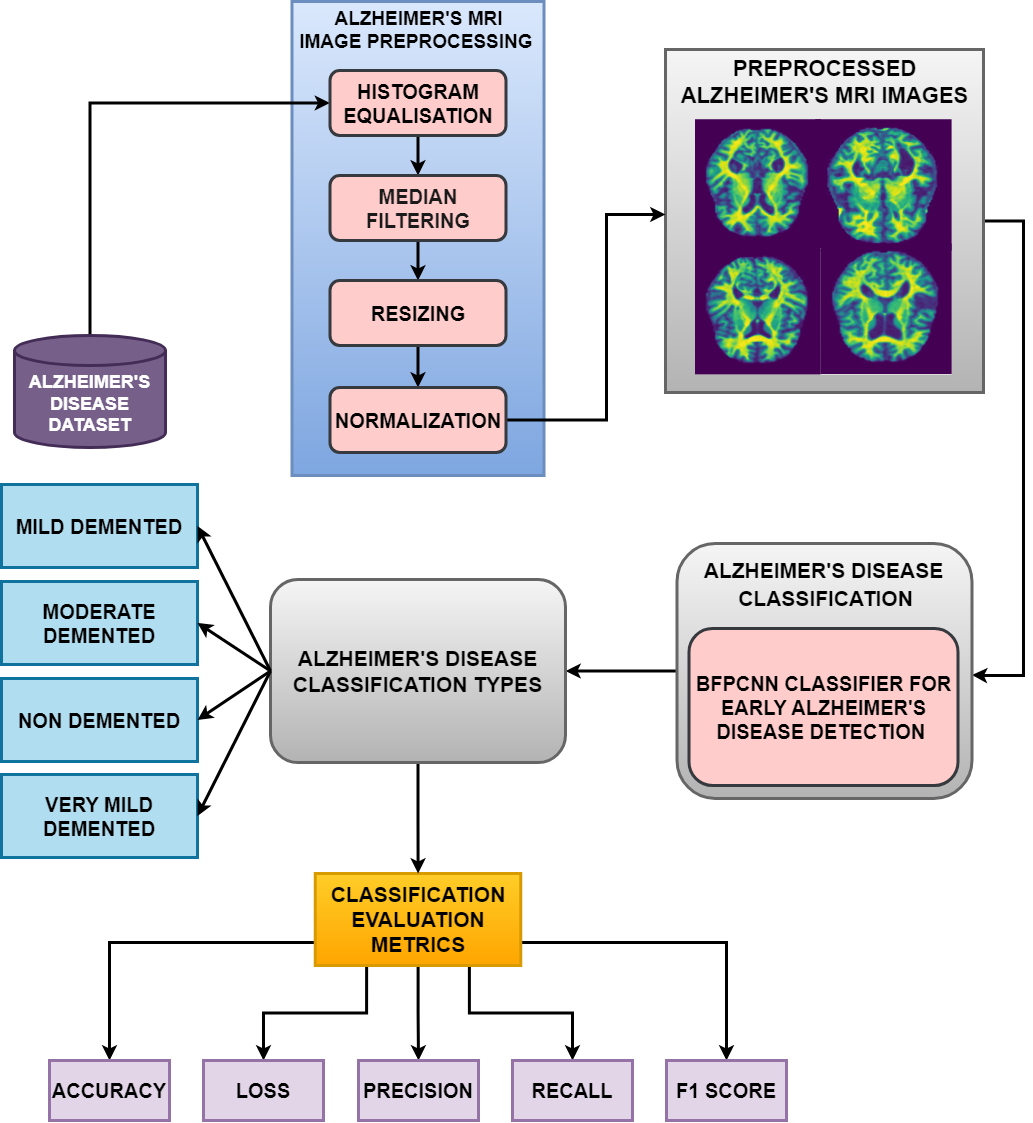}
  \caption{Overall Proposed System Workflow for Alzheimer’s Disease Classification}
  \label{fig:fig2}
\end{figure*}

We used the OASIS project to facilitate neuroimaging datasets of the brain in this study. The research utilized cross-sectional MRI data for the young and middle-aged individuals, old age non-demented, and demented subjects through the OASIS-1 dataset. This dataset comprises MRI scans of 416 male and female subjects who are aged between 18 and 96. For every subject it covers 3 to 4 T1-weighted-scans acquired in a single session. It is also made up of scans of subjects who are assessed to be not demented and those diagnosed with mild to moderate AD.

\begin{figure}
    \centering
    \includegraphics[width=\linewidth]{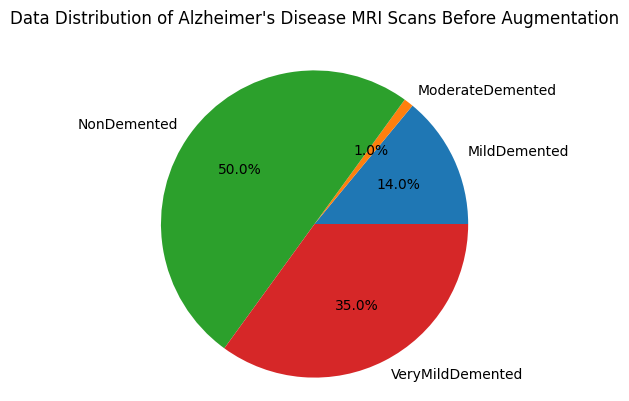}
    \caption{Graphical Representation of Data Distribution of Each Type of Alzheimer's Disease MRI Scans Before Augmentation}
    \label{fig:pi1}
\end{figure}

It had been noticed that the OASIS dataset was majorly imbalanced in terms of classes as has been projected in Table 1. The MRI scans under various types of AD, such as 'MildDemented', 'ModerateDemented', 'NonDemented', 'VeryMildDemented', have been tabulated in Table \ref{tab:table1} and graphically represented in Figure \ref{fig:pi1}.

\begin{table}[h]
  \centering
  \caption{Image Count of Each Type of Alzheimer's Disease MRI Scans Before Augmentation}
   \vspace{-5pt}
  \begin{tabular}{@{}ll@{}} 
    \toprule
    \textbf{Type of Alzheimer’s Disease} & \textbf{Images Count} \\
    \midrule
    Non Demented          & 2560 \\
    Very Mild Demented    & 1792 \\
    Mild Demented         & 717  \\
    Moderate Demented     & 52   \\
    \bottomrule
  \end{tabular}
  \label{tab:table1}
\end{table}

\begin{figure}
    \centering
    \includegraphics[width=\linewidth]{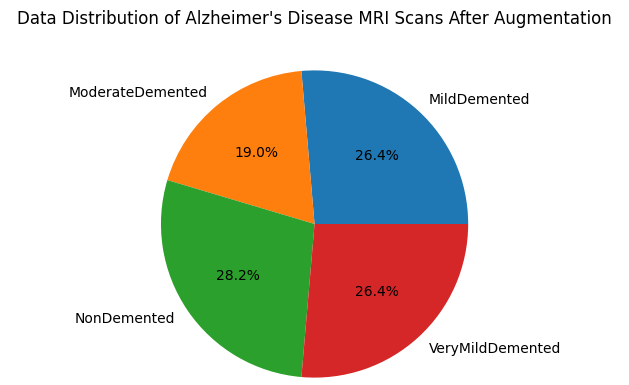}
    \caption{Graphical Representation of Data Distribution of Each Type of Alzheimer's Disease MRI Scans After Augmentation}
    \label{fig:pi2}
\end{figure}

Balancing this is crucial for guaranteeing that DL models are learned properly. To that effect, the Augmented Alzheimer MRI Dataset \cite{uraninjo_augmented_alzheimer_mri} was used, which contains augmented images for every class of Alzheimer's MRI scans. Since augmentation balances the image distribution among all classes, it resolved the issue of class imbalance for this data set. The images' distribution for each type of AD has been presented in Table \ref{tab:table2} and graphically represented in Figure \ref{fig:pi2}.

\begin{table}[h]
 \caption{Image Count of Each Type of Alzheimer's Disease MRI Scans After Augmentation}
  \centering
  \vspace{-5pt}
  \begin{tabular}{ll}
    \toprule
    \textbf{Type of Alzheimer’s Disease} & \textbf{Images Count} \\
    \midrule
    Non Demented                & 9600         \\
    Very Mild Demented          & 8960         \\
    Mild Demented               & 8960         \\
    Moderate Demented           & 6464         \\
    \bottomrule
  \end{tabular}
  \label{tab:table2}
\end{table}

\subsection{Methodology}

We first preprocessed the augmented MRI images using four preprocessing methods, namely histogram equalization, median filtering, image resizing, and pixel normalization. Histogram equalization is employed to improve the contrast of the images. Median filtering reduces the noise within an image but does not blur the edges of an important structure. Resizing the dimension of all images passed to the model ensures that it receives images of equal dimensions. Normalization contributes to fast training of the model by eliminating unacceptably large ranges of input values, which might slow down learning. We then used the above-preprocessed images to train our proposed classifier model and classify the MRI images among one of the four AD types. We then evaluated the classification results using metrics like F1-score, recall, accuracy, and precision. The flow of work has been depicted in Figure \ref{fig:fig2}.

\subsection{Preprocessing}
\begin{figure*}[h]
  \centering
  \includegraphics[width=0.7\textwidth]{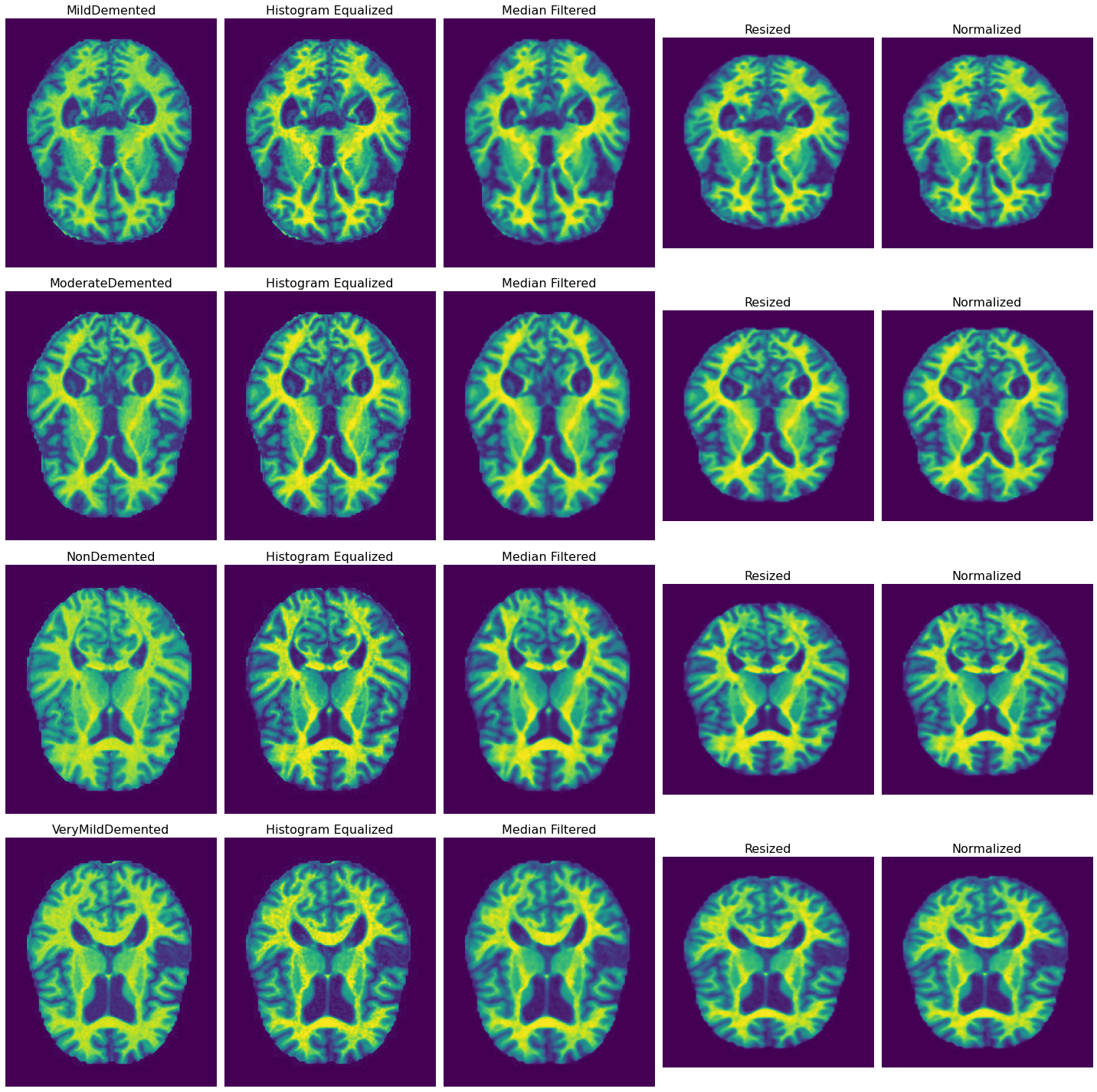}
  \caption{Alzheimer’s Disease MRI Scans Observed After Each Pre-processing Step}
  \label{fig:figPrepAll}
\end{figure*}

We preprocess the augmented images in the dataset by employing a pipeline that includes four stages: histogram equalization, median filtering, image resizing, and pixel normalization. The results following each preprocessing step of the Alzheimer's MRI image have been shown in Figure \ref{fig:figPrepAll}.

\subsubsection{Histogram Equalization}
Histogram Equalization is used for enhancing the contrast of an image. It redistributes pixel intensity levels. These are made to overcome difficulties of distinguishing among several tissues or structures in MRI due to the wide range of intensity values arising in many of these MRI scans. Histogram equalization stretches the intensity values of the single pixels across the possible range of intensity values.
\begin{equation}
\underbrace{I_{equalized}(x,y)}_{\text{Equalized Image}} = \underbrace{\left(\frac{L-1}{M \cdot N}\right)}_{\text{Scaling Factor}} \cdot \underbrace{\sum_{k=0}^{I_{in}(x,y)} \underbrace{H(k)}_{\text{Histogram}}}_{\text{Cumulative Distribution Function}}
\end{equation}
This preprocessing technique ensures that the contrast between different tissues or structures in an MRI scan is enhanced with a view of identification and analysis. This helps where the original MRI scan has poor contrast, or some features are not distinguishing. The algorithmic workflow of this step is presented in Algorithm 1.

\noindent
\begin{table}[h!]
\centering
\renewcommand{\arraystretch}{1.2}
\setlength{\tabcolsep}{5pt}
\begin{tabular}{@{}p{0.9\columnwidth}@{}}
\toprule
\textbf{Algorithm 1: Histogram Equalization} \\ 
\midrule
\textbf{Input:} raw MRI image \\
\textbf{Output:} histogram equalized MRI image \\ 
\textbf{Function} histogramEqualization(image$_{\text{MRI}}$): \\
1. $hist_{image} \leftarrow$ histogram of image$_{\text{MRI}}$ \\
2. Compute cumulative distribution from $hist_{image}$: \\
\hspace{1cm} $CDF(i) \leftarrow \sum_{j=0}^i P_r(j)$ \\
3. Transform pixel intensity $I(x,y)$ at coordinate $(x,y)$ using $CDF$: \\
\hspace{1cm} $I_{\text{equalised}}(x,y) \leftarrow \text{round}\left(\frac{CDF(I(x,y)) - CDF_{\text{min}}}{CDF_{\text{max}} - CDF_{\text{min}}} \cdot (L-1)\right)$ \\
4. \textbf{For} each pixel $p$ in image$_{\text{MRI}}$ \textbf{do} \\
\hspace{1cm} Apply histogram equalization transformation on pixel $p$ \\
5. \textbf{End For} \\
6. image$_{\text{heq}} \leftarrow$ histogram equalized image \\
7. \textbf{Return} image$_{\text{heq}}$ \\
\bottomrule
\end{tabular}
\label{tab:alg1}
\end{table}

\subsubsection{Median Filtering}
AD MRI scans are susceptible to boundary effects and hence MRI scans tend to capture noise, such as Gaussian noise or salt-and-pepper noise, which degrades the image quality. Median filtering replaces every pixel by the median value of surrounding pixels, thereby preserving the edges and details of structures in the image and effectively removing noise-induced outliers. It will permit better general quality MRI scans for AD, which also means better extraction of features. The algorithmic workflow of median filtering is presented in Algorithm 2.
\begin{equation}
\begin{split}
\underbrace{I_{out}(a,b)}_{\text{Output Image}} = 
\underbrace{\text{median}}_{\text{Median Operator}} \Big(
\underbrace{\{I_{in}(a+m,b+n) \mid m, n \in W\}}_{\text{Neighborhood Pixels}}
\Big)
\end{split}
\end{equation}

\noindent
\begin{table}[h!]
\centering
\renewcommand{\arraystretch}{1.2}
\setlength{\tabcolsep}{5pt}
\begin{tabular}{@{}p{0.9\columnwidth}@{}}
\toprule
\textbf{Algorithm 2: Median Filtering} \\ 
\midrule
\textbf{Input:} histogram equalized MRI image \\
\textbf{Output:} median filtered MRI image \\ 
\textbf{Function} median\_filter(image$_{\text{heq}}$): \\
1. $image \leftarrow image_{\text{heq}}$ \\
2. $l \leftarrow$ length of image \\
3. $b \leftarrow$ breadth of image \\
4. $c \leftarrow$ channels of image \\
5. $w \leftarrow 3$ \textbf{// window\_size} \\
6. $filtered\_image \leftarrow$ create\_empty\_image$(l, b)$ \\
7. $c_{\text{img}} \leftarrow$ image split into channel \\
8. \textbf{For} $i = 0$ to $l - 1$ \textbf{do} \\
\hspace{1cm} \textbf{For} $j = 0$ to $b - 1$ \textbf{do} \\
\hspace{2cm} $c_{\text{img}} += image[i][j][1]$ \\
\hspace{1cm} \textbf{End For} \\
8. \textbf{End For} \\
9. apply\_median\_filter($c_{\text{img}}, w$) \\
10. \textbf{For} $i = 0$ to $l - 1$ \textbf{do} \\
\hspace{1cm} \textbf{For} $j = 0$ to $b - 1$ \textbf{do} \\
\hspace{2cm} $image_{\text{filtered}} = [c_{\text{img}}[i][j]]$ \\
\hspace{1cm} \textbf{End For} \\
10. \textbf{End For} \\
11. \textbf{Return} $image_{\text{filtered}}$ \\
\textbf{End Function} \\
\bottomrule
\end{tabular}
\label{tab:alg2}
\end{table}

\newpage

\subsubsection{MRI Image Resizing}
Resizing is used to prevent overfitting in the model since the input data reduced its complexity. DL models must be a fixed size input; that is to say, all images presented to the model are the same size, which will be particularly important for batch processing, and it is consistent throughout the dataset. It also because resizing would take less time to train the model, providing better efficiency in model training. The algorithmic workflow of this preprocessing step is presented in Algorithm 3.
\begin{equation}
\underbrace{I_{out}(u,v)}_{\text{Resized Image}} = \underbrace{\sum_{i=0}^{M-1} \sum_{j=0}^{N-1} \underbrace{I_{in}(i,j)}_{\text{Input Image}} \cdot \underbrace{K(u - S_u(i), v - S_v(j))}_{\text{Interpolation Kernel}}}_{\text{Interpolation Sum}}
\end{equation}

\noindent
\begin{table}[h!]
\centering
\renewcommand{\arraystretch}{1.2}
\setlength{\tabcolsep}{5pt}
\begin{tabular}{@{}p{0.9\columnwidth}@{}}
\toprule
\textbf{Algorithm 3: Image Resizing} \\ 
\midrule
\textbf{Input:} median filtered MRI image \\
\textbf{Output:} resized MRI image \\ 
\textbf{Function} resize(image$_{\text{filtered}}$): \\
1. $h \leftarrow$ height of image$_{\text{filtered}}$ \\
2. $w \leftarrow$ width of image$_{\text{filtered}}$ \\
3. $t \leftarrow$ target size \\
4. Calculate scaling factor: \\
\hspace{1cm} $S_H \leftarrow \frac{t}{h}$ \\
\hspace{1cm} $S_W \leftarrow \frac{t}{w}$ \\
5. \textbf{For each} pixel at coordinates $(x, y)$ \textbf{do} \\
\hspace{1cm} $i' \leftarrow \lfloor x \cdot S_H \rfloor$ \\
\hspace{1cm} $j' \leftarrow \lfloor y \cdot S_W \rfloor$ \\
\hspace{1cm} $image_{\text{resized}}(i', j') \leftarrow image_{\text{MRI}}(i, j)$ \\
6. \textbf{End For} \\
7. \textbf{Return} $image_{\text{resized}}$ \\
\textbf{End Function} \\
\bottomrule
\end{tabular}
\label{tab:alg3}
\end{table}

\subsubsection{Pixel Normalization}
Pixel Normalization normalizes pixel values in the images between the range (0, 1). It reduces large ranges of input values that can impair learning and brings the model to convergence faster in training. After normalizing input images, gradients tend to be more stable and less prone to getting vanishing effects in the gradient descent optimization algorithms applied in DL models. This also prevents overfitting since it makes sure that the model learns the patterns of the data and does not memorize the pixel values. The algorithmic workflow of pixel normalization is presented in Algorithm 4.
\begin{equation}
\underbrace{I_{norm}(i,j)}_{\text{Normalized Pixel}} = \frac{\underbrace{I(i,j)}_{\text{Pixel Value}}}{\underbrace{255.0}_{\text{Maximum Pixel Value}}}
\end{equation}
\begin{table}[h!]
\centering
\renewcommand{\arraystretch}{1.2}
\setlength{\tabcolsep}{5pt}
\begin{tabular}{@{}p{0.9\columnwidth}@{}}
\toprule
\textbf{Algorithm 4: Normalization} \\ 
\midrule
\textbf{Input:} resized MRI image \\
\textbf{Output:} normalized MRI image \\ 
\textbf{Function} normalize\_image(image$_{\text{resized}}$): \\
1. $l \leftarrow$ length of image$_{\text{resized}}$ \\
2. $b \leftarrow$ breadth of image$_{\text{resized}}$ \\
3. $c \leftarrow$ channels of image$_{\text{resized}}$ \\
4. $normalisation\_value \leftarrow 255$ \\
5. $normalised\_image \leftarrow$ create\_empty\_image$(l, b)$ \\
6. \textbf{For} $i = 0$ to $l - 1$ \textbf{do} \\
\hspace{0.75cm} \textbf{For} $j = 0$ to $b - 1$ \textbf{do}\\
\hspace{1.35cm} \textbf{For} $k = 0$ to $c - 1$ \textbf{do}\\
\hspace{1.75cm} \resizebox{!}{\baselineskip}{$
normalised\_image[i][j][k] \leftarrow \frac{image_{\text{resized}}[i][j][k]}{255}
$}\\
\hspace{1.35cm} \textbf{End For}\\
\hspace{0.75cm} \textbf{End For}\\
6. \textbf{End For} \\
7. \textbf{Return} $normalised\_image$ \\
\textbf{End Function} \\
\bottomrule
\end{tabular}
\label{tab:alg4}
\end{table}

\subsection{Granular Feature Integration}

Detecting Alzheimer’s Disease (AD) through MRI scans demands models capable of extracting hierarchical features \( \mathcal{F} = \{ f_1, f_2, \dots, f_n \} \) that span both macro- and micro-level brain structures. Conventional Convolutional Neural Networks (CNNs), formulated as \( f_{\theta}(x) = \sigma(W * x + b) \) with convolutional weights \( W \) and biases \( b \), often excel at low-level feature extraction but struggle with high-level granular details critical for AD detection. To overcome this, we propose Granular Feature Integration (GFI), a mechanism that fuses residual learning \( R(x) = x + \mathcal{H}(x) \) with multi-scale processing \( M(x) = \bigoplus_{s \in S} \psi_s(x) \), where \( S \) denotes the set of scales and \( \psi_s \) represents convolutional transformations at scale \( s \).

Formally, GFI can be represented as a composite function:
\begin{equation}
GFI(x) = R(I(x)),
\end{equation}
where \( R(x) = x + F(x, W) \) denotes the residual mapping and \( I(x) = \bigoplus_{k=1}^{n} \delta_k(x) \) aggregates multi-scale features through convolutions \( \delta_k \) with kernel sizes \( k \in \{1, 3, 5, 7\} \). The residual pathways ensure that the gradient \( \nabla_{x} L \) remains bounded, thus solving the vanishing gradient problem for deep architectures, while the inception modules allow the network to process spatial hierarchies by varying the receptive field \( r_k \).

By integrating these two mechanisms, the model captures a more expressive feature space:
\begin{equation}
\mathcal{F}_{\text{granular}} = \{ f_{\text{global}}, f_{\text{local}} \},
\end{equation}
where \( f_{\text{global}} \) represents macroscopic brain structures, and \( f_{\text{local}} \) encodes subtle neurodegenerative markers.

\subsection{Proposed Bi-Focal Perspective Convolutional Neural Network (BFPCNN)}
\begin{figure*}[h]
  \centering
  \includegraphics[width=\textwidth]{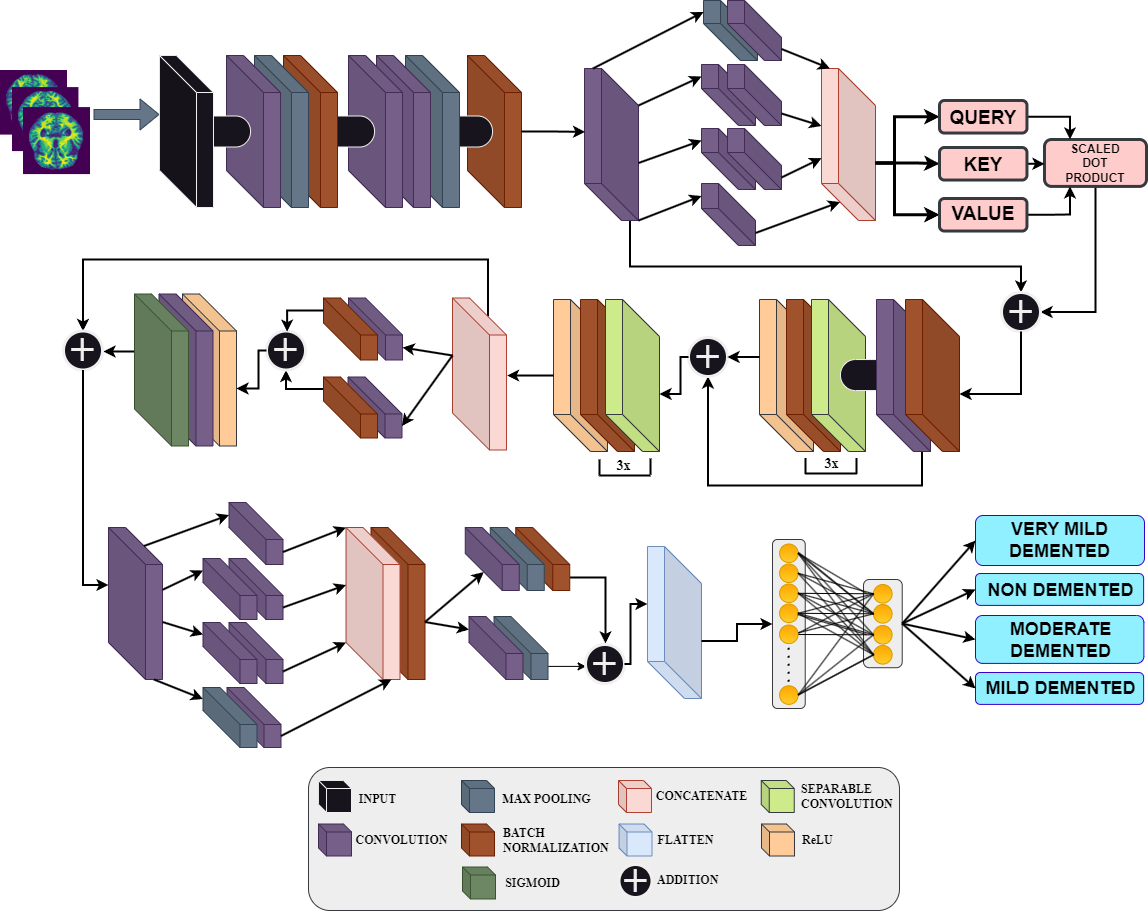}
  \caption{Proposed BFPCNN Layer Architecture for AD Classification}
  \label{fig:fig4}
\end{figure*}

In this section, we present our proposed Bi-Focal Perspective CNN for Early Alzheimer's Detection. We feed the preprocessed MRI images to our proposed model, and then our model is going to output the respective predicted probabilities for each disease type. The overall architecture of the proposed BFPCNN is shown in Figure \ref{fig:fig4}. 

To get attention to detail in MRI scans along with eliminating vanishing gradients along layers, our model has focused on a Bi-Focal Perspective mechanism. For efficiently extracting fine details and raising feature representation, we have incorporated Granular Feature Integration to improve the overall capability of the model to capture features of different scales. It has the ability to correctly lift out the desired features and classify very well among the varieties of AD, thus improving the performance significantly.
Our model begins with an initial convolutional block, consisting of a convolution layer that utilizes 64 filters and a kernel of size 7x7, followed by a max-pooling layer. The basic features the model identifies include the layout of the brain and contours of various brain regions from the MRI image. Then comes the next convolutional block, which consists of a Batch Normalization layer and further convolutional layers with heterogeneous parameters. This block helps the model to refine the initial features and in the identification of gray and white matter, which is basic for recognizing anatomical features in the brain.

\begin{equation}
\underbrace{\text{Conv}(I_{(i,j)}, F)}_{\text{Convolution Operation}} = \sum_{m=0}^{M-1} \sum_{n=0}^{N-1} \underbrace{I_{(i+m,j+n)}}_{\text{Input feature map}} \cdot \underbrace{F_{(m,n)}}_{\text{Filter}} + \underbrace{b}_{\text{Bias}}
\end{equation}

\begin{equation}
\underbrace{\text{MaxPooling}(O)_{(i,j)}}_{\text{Max Pool Operation}} = \max_{p=0}^{k-1} \max_{q=0}^{k-1} \underbrace{I_{(i \cdot s + p, j \cdot s + q)}}_{\text{Input feature map}}
\end{equation}

\begin{equation}
\begin{split}
    \underbrace{\text{BN}(x)}_{\text{Batch Normalization}} = & \underbrace{\gamma}_{\text{Adaptable parameter}} \left( \frac{\underbrace{x}_{\text{Input}} - \underbrace{\mu}_{\text{Mean}}}{\sqrt{\underbrace{\sigma^2}_{\text{Variance}} + \underbrace{\epsilon}_{\text{constant}}}} \right) + \\ 
    & \underbrace{\beta}_{\text{Adaptable parameter}}
\end{split}
\end{equation}

The feature map is then passed to an inception block \cite{szegedy2015going}. This allows expansion in the capacity of capturing various features at different scales. It consists of parallel paths to convolutional layers with varying kernel sizes (5x5, 1x1, and 3x3) and max pooling layers. This helps the model extract fine details while also focusing on broader contextual information simultaneously. Smaller-sized kernels look at the local features and try to monitor slight changes in the hippocampal region of the brain. Larger-size kernels look at the ventricles enlargement overall, which is believed to be a typical sign in later stages of Alzheimer's. These parallel paths then get concatenated to form one feature vector that comes from the inception block. The algorithmic workflow of inception block is presented in Algorithm 5.

\begin{equation}
\resizebox{\columnwidth}{!}{$
\begin{aligned}
\underbrace{\text{concat}(X, Y)_{(p,q,r)}}_{\text{concatenation operation}} = 
\begin{cases} 
\underbrace{X_{(p,q,r)}}_{\text{input feature map from X}} 
& \text{if } 1 \leq r \leq \text{depth}(X), \\[10pt]
\underbrace{Y_{(p,q,r-\text{depth}(X))}}_{\text{input feature map from Y}} 
& \text{if } \text{depth}(X) < r \leq \text{depth}(X) + \text{depth}(Y).
\end{cases}
\end{aligned}
$}
\end{equation}

\begin{table}[h!]
\centering
\renewcommand{\arraystretch}{1.2}
\setlength{\tabcolsep}{5pt}
\begin{tabular}{@{}p{0.9\columnwidth}@{}}
\toprule
\textbf{Algorithm 5: Inception Block} \\ 
\midrule
\textbf{Function} inceptionBlock(input): \\
1. $x_{11} \leftarrow \text{number of filters for path 1 conv}_{1}$ \\
2. $x_{21} \leftarrow \text{number of filters for path 2 conv}_{1}$ \\
3. $x_{22} \leftarrow \text{number of filters for path 2 conv}_{2}$ \\
4. $x_{31} \leftarrow \text{number of filters for path 3 conv}_{1}$ \\
5. $x_{32} \leftarrow \text{number of filters for path 3 conv}_{2}$ \\
6. $x_{41} \leftarrow \text{number of filters for path 4 conv}_{1}$ \\
7. $y \leftarrow kernel size$ \\
8. $z \leftarrow convolution stride$ \\
9. $p_{1} \leftarrow conv2D(x_{11}, y, z)(input)$ \\
10. $p_{2} \leftarrow conv2D(x_{21}, y, z)(input)$ \\
11. $p_{2} \leftarrow conv2D(x_{22}, y, z)(p2)$ \\
12. $p_{3} \leftarrow conv2D(x_{31}, y, z)(input)$ \\
13. $p_{3} \leftarrow conv2D(x_{32}, y, z)(p3)$ \\
14. $p_{4} \leftarrow maxPooling(2,2)(input)$ \\
15. $p_{4} \leftarrow conv2D(x_{41}, y, z)(p_{4})$ \\
16. $output \leftarrow concatenate(p_{1}, p_{2}, p_{3}, p_{1})(input)$ \\
17. \textbf{Return} $output$ \\
\textbf{End Function} \\
\bottomrule
\end{tabular}
\label{tab:alg4}
\end{table}

Following this initialization block, there is a self-attention mechanism \cite{vaswani2017attention} to determine the importance of features relative to one another. The models are, therefore, enabled to focus in these areas that often reflect typical features of AD like altered brain volume. The algorithmic workflow of self-attention is presented in Algorithm 6.

\begin{equation}
\resizebox{\columnwidth}{!}{$
\underbrace{\alpha_{ij}}_{\text{Attention weight for position } (i,j)} = \underbrace{\text{softmax}\left(\frac{\underbrace{K^TQ}_{\text{Dot product of key and query}}}{\sqrt{\underbrace{d_k}_{\text{Dimensionality of key vectors}}}}\right)}_{\text{Softmax normalization}} \underbrace{V}_{\text{Value matrix}}
$}
\end{equation}

\begin{table}[h!]
\centering
\renewcommand{\arraystretch}{1.2}
\setlength{\tabcolsep}{5pt}
\begin{tabular}{@{}p{0.9\columnwidth}@{}}
\toprule
\textbf{Algorithm 6: Self Attention} \\ 
\midrule
\textbf{Function} selfAttention(input): \\
1. $q \leftarrow \text{query vector}$ \\
2. $k \leftarrow \text{key vector}$ \\
3. $v \leftarrow \text{value vector}$ \\
4. $s \leftarrow \text{scaling parameter}$ \\
5. $a \leftarrow (q@k.transpose(-2,-1)) * s$ \\
6. $matrix_{qk} \leftarrow softmax(a) $\\
7. $z \leftarrow Dropout(a) $\\
8. $z \leftarrow fullyConnected(z) $\\
9. $z \leftarrow Dropout(z) $\\
10. \textbf{Return} $z.matrix_{qk}$ \\
\textbf{End Function} \\
\bottomrule
\end{tabular}
\label{tab:alg4}
\end{table}

This feature map then passes through separable convolutional blocks with multiple residual connections \cite{he2016deep} ensuring proper gradients flow through the neural network. Separable convolutional layers split the convolutions into depthwise and pointwise parts, which ensure minimal computational complexity while retaining the capability to capture intricate spatial dependencies from the MRI scan. Batch normalization normalizes the activations of the layer, ensuring that the learning is stabilized and the internal covariate shifts are reduced. ReLU activation functions induce a nonlinearity in the models. This is necessary for extracting more complex relationships from the features extracted.

\begin{equation}
\underbrace{\text{ReLU}(x)}_{\text{Rectified Linear Unit Activation}} =
\begin{cases}
\underbrace{x}_{\text{Feature map}} & \text{if } x > 0 \\
0 & \text{otherwise}
\end{cases}
\end{equation}


After the separable convolutional blocks, a spatial attention mechanism \cite{woo2018cbam} is applied. It emphasizes the crucial regions of interest in the spatial domain of the feature map. These uses of both spatial attention as well as self-attention, coming under the Bi-Focal Perspective, help the model pay attention to local and global features. As spatial attention pays attention to the spatial domain, self-attention does its work in enabling the model to analyze the association between features across the full feature map and capture subtle changes for analysis. The algorithmic workflow of spatial-attention is presented in Algorithm 3. The combined application of both mechanisms helps the Bi-Focal Perspective aid the model in understanding the subtle details as well as developing an understanding of the bigger picture in the MRI scan.

\vspace{-12pt}
\noindent
\begin{table}[h!]
\centering
\renewcommand{\arraystretch}{1.2}
\setlength{\tabcolsep}{5pt}
\begin{tabular}{@{}p{0.9\columnwidth}@{}}
\toprule
\textbf{Algorithm 7: Spatial Attention} \\ 
\midrule
\textbf{Function} spatialAttention(input): \\
1. $d \leftarrow$ dilation convolution count \\
2. $x \leftarrow$ number of filters for conv \\
3. $y \leftarrow$ kernel size \\
4. $z \leftarrow$ convolution stride \\
5. \textbf{For} $i = 1$ to $d$ \textbf{do} \\
\hspace{1cm} $t_i \leftarrow \text{conv}(f, k, s)(\text{input})$ \\
\hspace{1cm} $t_i \leftarrow \text{batchNorm}(t_1)$ \\
6. \textbf{End For} \\
7. $output \leftarrow \sum_i t_i \quad \forall t \in [1, 2, \dots, d]$ \\
8. \textbf{Return} $output$ \\
\textbf{End Function} \\
\bottomrule
\end{tabular}
\end{table}

This is further followed by a second inception block that complements the spatial attention mechanism by providing an overall view of the MRI data at different levels. Then for even better learning of residual functions, a residual block has been implemented. Then there exists a flattening layer used for flattening down the output to a dimension for dense processing in a dense layer. Finally, a softmax activation function returns a probability distribution of the input MRI image over the different classes – 'Mild Demented', 'Moderate Demented', 'Non-Demented', and 'Very Mild Demented'.

\begin{equation}
\underbrace{S(I_i)}_{\text{Softmax Output}} = \frac{\underbrace{e^{I_i}}_{\text{Exponentiated Input}}}{\underbrace{\sum_{j=1}^N e^{I_j}}_{\text{Sum of Exponentiated Inputs}}}
\end{equation}

The parameter specifications of the proposed BFPCNN are presented in Table \ref{tab:tab3}.

\begin{table}[h]
 \caption{Proposed Model Parameter Specifications}
  \centering
  \vspace{-5pt}
  \begin{tabular}{ll}
    \toprule
    \textbf{Parameters}             & \textbf{Coefficients} \\
    \midrule
    Total number of layers & 68           \\
    Learning rate          & 0.001        \\
    Epochs                 & 100          \\
    Batch size             & 128          \\
    Total Parameters       & 153,066,236  \\
    \bottomrule
  \end{tabular}
  \label{tab:tab3}
\end{table}

\section{Results and Discussion}
\subsection{Experimental Setup}

In this section, we discuss the performance of the proposed model by comparing it with other models in the same domain. We carried out our experiments on a system characterized by the following specifications: Operating System - Linux 5.15.133; CPU - AMD EPYC 7763 with 128 CPU(s) and x86\_64 architecture; GPU - AMD Radeon Instinct Model 1; with 64 Core(s) per socket, 2 Socket(s), and 1 Thread(s) per core. This information is summarized in Table 4.

\begin{table}[h]
 \caption{Configuration and Specifications of the System Used}
  \centering
  \vspace{-5pt}
  
  \begin{tabular}{ll}
    \toprule
    \textbf{Component}& \textbf{Specification}         \\
    \midrule
    Operating System                  & Linux 5.15.133                 \\
    CPU                               & AMD EPYC 7763                  \\
    Architecture                      & x86\_64                        \\
    CPU(s)                            & 128                            \\
    GPU                               & AMD Radeon Instinct            \\
    Model                             & 1                              \\
    Core(s) per socket                & 64                             \\
    Socket(s)                         & 2                              \\  
    Thread(s) per core                & 1                              \\
    \bottomrule
  \end{tabular}
  \label{tab:system_specifications}
\end{table}

\subsection{Evaluation Metrics}
Our proposed model’s performance for detecting AD was evaluated using the following metrics.
\subsubsection{Accuracy}
Accuracy can be defined as the evaluation metric which is responsible for determining how close the predicted value is to its true value. It is computed as a ratio between the count of true positives to the count of all the samples given to the model.
\begin{equation}
\underbrace{Accuracy}_{\text{Accuracy}} = \frac{\underbrace{\sum_{i=1}^N \underbrace{TP_i}_{\text{TruePositives}} + \underbrace{TN_i}_{\text{TrueNegatives}}}_{\text{CorrectPredictions}}}{\underbrace{N}_{\text{TotalSamples}}}
\end{equation}
Where n represents the number of classes present in the dataset, TP      represents the number of true positive samples, and FP,     FN,      and TN       represent the number of false positive, false negative and true negative samples respectively. 

\subsubsection{Precision}
To measure the model’s performance which involves the identification ability of only the relevant instances, the precision metric is taken into account and is considered critical when there is an increase in the number of false positives. It denotes the ratio of true positives to the sum of true positives and false positives.
\begin{equation}
\underbrace{Recall_c}_{\text{Recall for Class } c} = \frac{\underbrace{TP_c}_{\text{True Positives for Class } c}}{\underbrace{TP_c}_{\text{True Positives for Class } c} + \underbrace{FN_c}_{\text{False Negatives for Class } c}}
\end{equation}

\subsubsection{Recall}
To measure the model’s performance which involves the capture ability of all the relevant instances, the recall metric is taken into account and is considered critical when there is an increase in the number of false negatives. It represents the proportion of true positives relative to the sum of true positives and false negatives.
\begin{equation}
\underbrace{Precision_c}_{\text{Precision for Class } c} = \frac{\underbrace{TP_c}_{\text{True Positives for Class } c}}{\underbrace{TP_c}_{\text{True Positives for Class } c} + \underbrace{FP_c}_{\text{False Positives for Class } c}}
\end{equation}

\subsubsection{F1-Score}
The F1-Score integrates precision and recall into a unified metric, providing a balanced assessment of the model’s performance. It is calculated as the harmonic mean of precision and recall.
\begin{equation}
\underbrace{\text{F1 Score}}_{\text{F1 Score for Class c}} = 2 \cdot \frac{\text{recall} \cdot \text{precision}}{\text{recall} + \text{precision}}
\end{equation}

\begin{figure}[ht]
    \centering
    \includegraphics[width=1\linewidth]{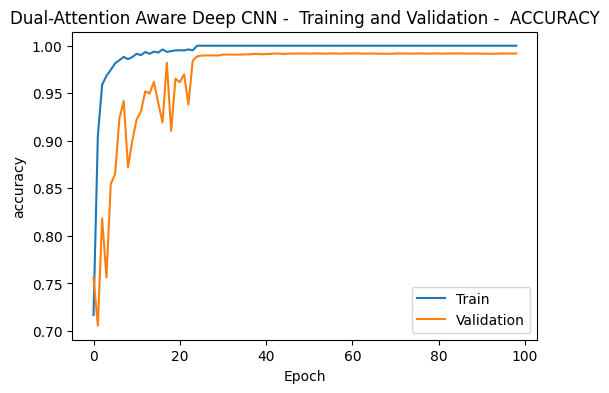}
    \caption{Training and validation accuracy plots for the Proposed Model}
    \label{fig:acc}
\end{figure}

\begin{figure}[ht]
    \centering
    \includegraphics[width=1\linewidth]{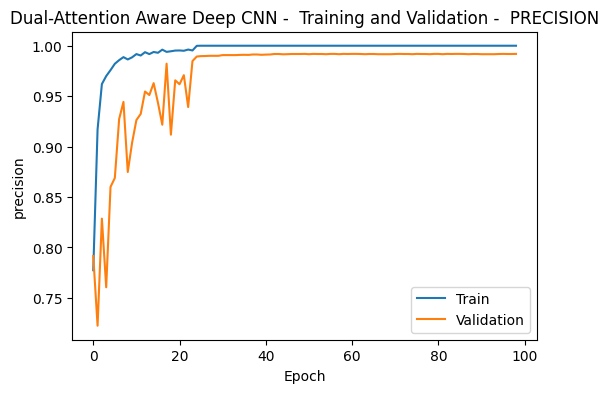}
    \caption{Training and validation precision plots for the Proposed Model}
    \label{fig:prec}
\end{figure}

\begin{figure}[ht]
    \centering
    \includegraphics[width=1\linewidth]{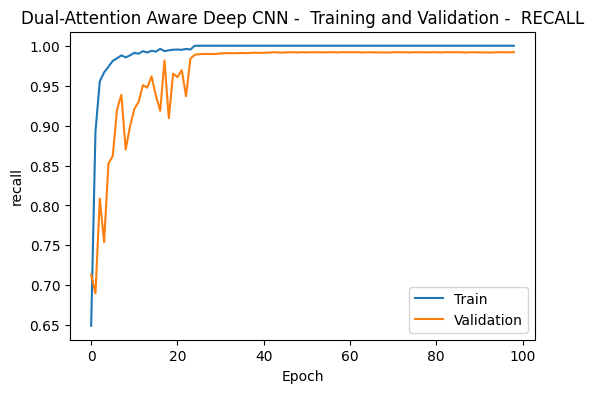}
    \caption{Training and validation recall plots for the Proposed Model}
    \label{fig:rec}
\end{figure}

\begin{figure}[ht]
    \centering
    \includegraphics[width=1\linewidth]{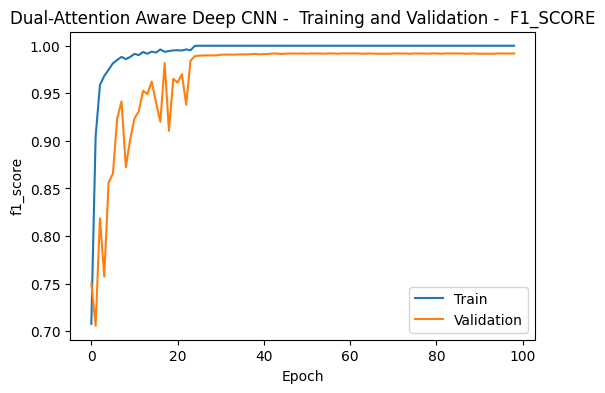}
    \caption{Training and validation F1-score plots for the Proposed Model}
    \label{fig:f1}
\end{figure}

\begin{figure}[ht]
    \centering
    \includegraphics[width=1\linewidth]{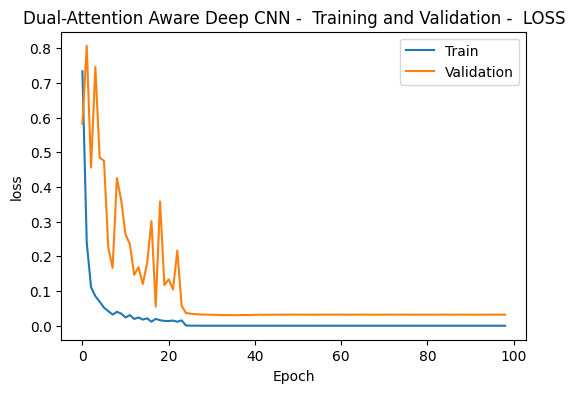}
    \caption{Training and validation loss plots for the Proposed Model}
    \label{fig:loss}
\end{figure}

Our classifier was trained for 100 epochs, and the performance metrics obtained at every epoch are recorded and plotted against each epoch to observe the variation in the model performance compared to its previous iterations. The scores obtained for accuracy, precision, recall and F1-score are illustrated in Figures \ref{fig:acc}, \ref{fig:prec}, \ref{fig:rec}, and \ref{fig:f1} respectively. The categorical cross entropy loss function was utilized to determine the loss incurred during training, and the obtained loss at each epoch is illustrated in Figure \ref{fig:loss}.

The above graphs show that our proposed model has demonstrated robust and exceptional performance during its learning phase. The model learns most of the features effectively up to around 30 epochs, after which it starts to converge. The proposed model achieved a precision of 99.3\%, a recall of 99.6\%, an F1-score of 99.3\%, and an accuracy of 99.2\%. These metrics indicate that the proposed model has learned the data well while being able to apply it to unseen scenarios effectively as seen from the validation curve. The Bi-Focal Perspective and Granular Feature Integration mechanisms integrated into the model allow it to effectively extract spatial dependencies from the MRI images, while also being able to apply more focus to specific parts of the image by comparing various aspects of the same image and ensuring proper flow of gradients along the network.

\subsection{Discussions}
The confusion matrix for the testing data using our trained proposed model is shown in Figure 10. This type of matrix displays values as proportions or percentages, making it easy to compare classification performance across different classes. With values ranging from 0 to 1, it simplifies interpretation. The matrix illustrates the classifier’s performance for the four AD classes: "Mild Demented," "Moderate Demented," "Non-Demented," and "Very Mild Demented." Each column represents the predicted class, while each row represents the actual (True) class. The main diagonal values indicate the percentage of correctly classified cases, whereas the off-diagonal values represent misclassified cases.

\begin{figure}
    \centering
    \includegraphics[width=1\linewidth]{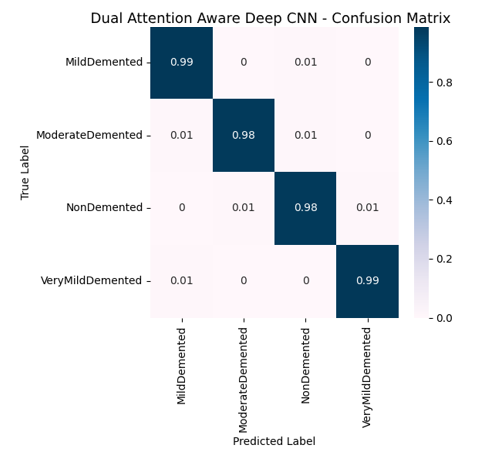}
    \caption{Confusion Matrix Obtained from the Proposed Model}
    \label{fig:enter-label}
\end{figure}

It can be observed from the confusion matrix, illustrated in Figure 10,  that our model has impressive accuracy across all classes. The majority of samples were correctly classified as shown by values close to one along the diagonal. For instance, for “Mild Demented” class, the model yielded a 0.99 true positive rate (TPR), implying that 99\% of samples were correctly classified. Again, for “Moderate Demented” class, had a TPR of 0.98 showing that 98\% of cases were correctly classified. The ‘Non-Demented’ class also had a high TPR of 0.98 expressing correct classification at 98\%. Lastly, the “Very Mild Demented” class had a 0.99 TPR, indicating that there was an accurate identification of about 99\% of samples in this category. In conclusion, these levels of misclassification rates indicated how effective these models are in differentiating various stages of Alzheimer’s ailment because they have small values off-diagonal.

Our proposed model was evaluated against other SOTA CNNs - DenseNet121 \cite{huang2016densely}, EfficientNetV2 \cite{tan2021efficientnetv2}, InceptionV3 \cite{szegedy2015going}, InceptionResNetV2 \cite{szegedy2016inception}, MobileNetV2 \cite{sandler2018inverted}, MobileNet \cite{howard2017mobilenets}, NASNetMobile \cite{zoph2017learning}, ResNet50 \cite{he2016deep}, VGG16 \cite{simonyan2015very}, and Xception \cite{chollet2016xception}. The comparison aimed to assess our model against established CNN benchmarks used as baselines for AD detection. The models were trained using the augmented OASIS dataset, and the results have been tabulated in Table 5. Evaluation of these models was based on key metrics such as Precision (Prec), Accuracy (Acc), F1-score (F1), and Recall (Rec).

\begin{table*}[htbp]
\centering
\caption{Proposed model performance comparison against pretrained CNNs during Training and Validation Phase}
\label{tab:table5}
\renewcommand{\arraystretch}{1.3} 
\setlength{\tabcolsep}{12pt}      

\Large

\resizebox{\textwidth}{!}{%
\begin{tabular}{lcccccccc}
\toprule
\textbf{Model} & \multicolumn{4}{c}{\textbf{Training Phase Metrics}} & \multicolumn{4}{c}{\textbf{Validation Phase Metrics}} \\ 
\cmidrule(lr){2-5} \cmidrule(lr){6-9}
               & \textbf{Accuracy} & \textbf{Precision} & \textbf{Recall} & \textbf{F1} & \textbf{Accuracy} & \textbf{Precision} & \textbf{Recall} & \textbf{F1} \\ 
\midrule
DenseNet121         & 0.9912 & 0.9913 & 0.9921 & 0.9914 & 0.9513 & 0.9622 & 0.9614 & 0.9625 \\
EfficientNetV2      & 0.9901 & 0.9912 & 0.9922 & 0.9913 & 0.9312 & 0.9423 & 0.9312 & 0.9311 \\
InceptionV3         & 0.9921 & 0.9911 & 0.9923 & 0.9912 & 0.9611 & 0.9613 & 0.9712 & 0.9613 \\
InceptionResNetV2   & 0.9512 & 0.9513 & 0.9623 & 0.9612 & 0.9323 & 0.9424 & 0.9313 & 0.9312 \\
MobileNetV2         & 0.9811 & 0.9822 & 0.9812 & 0.9823 & 0.8913 & 0.9014 & 0.9114 & 0.9013 \\
MobileNet           & 0.9712 & 0.9813 & 0.9723 & 0.9814 & 0.8914 & 0.9015 & 0.9013 & 0.9014 \\
NASNetMobile        & 0.9922 & 0.9923 & 0.9913 & 0.9924 & 0.8915 & 0.8814 & 0.8823 & 0.8824 \\
ResNet50            & 0.9912 & 0.9813 & 0.9813 & 0.9814 & 0.9213 & 0.9114 & 0.9114 & 0.9113 \\
VGG16               & 0.8813 & 0.8914 & 0.8814 & 0.8813 & 0.7814 & 0.7915 & 0.7814 & 0.7813 \\
Xception            & 0.9913 & 0.9914 & 0.9924 & 0.9913 & 0.9614 & 0.9523 & 0.9614 & 0.9613 \\
\textbf{Proposed CNN} & \textbf{0.9923} & \textbf{0.9914} & \textbf{0.9914} & \textbf{0.9924} & \textbf{0.9913} & \textbf{0.9923} & \textbf{0.9953} & \textbf{0.9933} \\
\bottomrule
\end{tabular}%
}
\end{table*}

From Table \ref{tab:table5}, we can infer that our proposed model has performed significantly well compared to other baseline CNN-based models. DenseNet121 benefits from dense connectivity, which aids in feature reusing; however, it impacts its ability to accurately identify particular Alzheimer’s features such as amyloid plaques and neurofibrillary tangles. EfficientNetV2 tends to overfit, due to challenges in capturing the subtle features of AD, such as the minor changes in brain structure or variations in brain activity observed in MRI scans. InceptionV3 is designed for efficient computation, making it incapable of analyzing the multidimensional characteristics of Alzheimer’s MRI data, especially in detecting subtle neural network disruptions or changes in brain volume. Despite its complicated structure, there is an inadequate match between InceptionResNetV2 and the demand to distinguish precise disease-specific features such as synaptic loss or alterations in brain function.

The lightweight design of MobileNetV2 is not as good at detailed analysis to identify conditions like the early symptoms of AD, that is, mild cognitive impairment (MCI) or minor hippocampal atrophy. By focusing on efficiency, MobileNet sacrifices the depth needed for thorough feature extraction from Alzheimer’s MRI images. The NASNetMobile, designed for mobile applications, is not fine-tuned enough to capture all the significant intersections that would signal Alzheimer’s such as altered white matter integrity or cortical atrophy. ResNet50’s depth and focus may make it difficult to learn about such things as hippocampal atrophy. Simplicity and the relative shallowness of VGG16 struggle with such detailed analysis where ventricular enlargement is observed resulting in low validation accuracy tests. Xception does not fully exploit the spatial relationships that are key to identifying specific indicators of Alzheimer’s, like brain atrophy patterns or connectivity between brain regions. Our proposed model stands out given that it combines dual spatial and self-attention mechanisms, providing a unique ability to identify amyloid deposition, neural network disruptions, and hippocampal atrophy among other critical features. These improvements assist in capturing intricate and heterogeneous pathology patterns typical for AD as seen from MRI scans.

Our model was also compared with existing research in the area of AD detection using ML and DL. The comparison of performance between these models is detailed in Table \ref{tab:table6}.

\begin{table}[htbp]
\centering
\caption{Proposed model performance metrics comparison with existing research}
\label{tab:table6}
\renewcommand{\arraystretch}{1.1} 
\setlength{\tabcolsep}{4pt}      

\footnotesize 

\resizebox{\columnwidth}{!}{%
\begin{tabular}{p{3.5cm}cccc}
\toprule
\textbf{Model} & \textbf{Accuracy (\%)} & \textbf{Recall (\%)} & \textbf{F1 (\%)} & \textbf{Precision (\%)} \\ 
\midrule
CNN, RNN, LSTM Ensemble \cite{dua2020cnnrnnlstm} & 89.75 & 91.07 & 89.08 & 89.16 \\
CNN, RNN, LSTM Bagged Ensemble \cite{dua2020cnnrnnlstm} & 92.22 & 92.57 & 91.87 & 91.92 \\
Wavelet Entropy, MLP and BBO \cite{wang2018single} & 92.41 & 92.47 & 92.30 & 92.14 \\ 
DEMNET without SMOTE \cite{murugan2021demnet} & 85.43 & 88.32 & 83.32 & 80.11 \\
DEMNET with SMOTE \cite{murugan2021demnet} & 95.23 & 95.21 & 95.27 & 96.43 \\ 
\textbf{Proposed Method} & \textbf{99.12} & \textbf{99.54} & \textbf{99.31} & \textbf{99.21} \\ 
\bottomrule
\end{tabular}%
}
\end{table}

From Table \ref{tab:table6}, we can infer that the CNN, RNN, and LSTM Ensemble model attained an accuracy of 89.75\%, with recall, precision, and F1 scores of 91.07\%, 89.16\%, and 89.08\%, respectively. The CNN, RNN, and LSTM Bagged Ensemble showed a slight improvement, with precision of 91.92\%, accuracy of 92.22\%, F1 score of 91.87\%, and recall of 92.57\%. The Wavelet Entropy, Multilayer Perceptron, and Biogeography-Based Optimization model performed comparably, achieving an accuracy of 92.4\%, precision of 92.14\%, recall of 92.47\%, and an F1 score of 92.3\%.
DEMNET without SMOTE had the lowest performance among the compared models, with a recall of 88.32\%, precision of 80.11\%, an F1 score of 83.32\%, and an accuracy of 85.43\%. However, when SMOTE was applied to DEMNET, the model’s performance significantly improved, achieving a recall of 95.21\%, precision of 96.43\%, an F1 score of 95.27\%, and an accuracy of 95.23\%,. Our proposed methodology demonstrates notable performance, with a recall of 99.54\%, precision of 99.21\%, F1 score of 99.31\%, and an accuracy of 99.12\%. This is because the model offers advanced feature extraction capabilities, and that this would help in identifying indicative images which would have features of hippocampal atrophy and cortical thinning. These enable the model to find very subtle and complex patterns and information associated with AD much more effectively.

\section{Conclusion and Future Scope}

AD is one among the slowly progressing and irreversible neurodegenerative disorders leading to serious complications like cell death in the neuronal structure. We hereby present a novel classification method for the diagnosis of AD using the application of Bi-Focal Perspectives and Granular Feature Integration as a promising approach toward early detection and a better diagnostics method. The high accuracy and precision scores of our proposed model indicate robustness and reliability in identifying and classifying AD. Graph Neural Networks, Graph Attention Networks, can therefore be explored for further improvement to the model's performance in future works. Capabilities of GATs and GNNs will help us capture the more complex relationships and dependencies within the neuroimaging data by utilizing both inter-regional and intra-regional connections. In addition, the models will be refined with more extensive and diverse datasets. Improved accuracy and generalization will be achieved as a result, promising better application in not only improving the diagnosis of Alzheimer's but also neurodegenerative disease research as a whole.


\bibliographystyle{IEEEtran}
\bibliography{assets/references}

\EOD

\end{document}